\documentclass[10pt, aps, prl, twocolumn, showpacs, showkeys, floatfix]{revtex4-1}
\usepackage{amsmath, amssymb, amsxtra, graphicx, array, mathtools, empheq, color, bbm}
\usepackage{hyperref}
\hypersetup{colorlinks=true, linkcolor=blue, citecolor=blue, urlcolor=blue}
\usepackage{times}

\newcommand{\trm}				{\textrm}

\begin{document}


\title{Quantum-coherent phase oscillations in synchronization}

\author{Talitha~Weiss}
\author{Stefan~Walter}
\author{Florian~Marquardt}
\affiliation{
Friedrich-Alexander University Erlangen-N{\"u}rnberg (FAU), Department of Physics, Staudtstra{\ss}e 7, 91058 Erlangen, Germany \\
and Max Planck Institute for the Science of Light, Staudtstra{\ss}e 2, 91058 Erlangen, Germany
}

\date{\today}

\pacs{05.45.Xt, 03.65.-w, 42.50.-p}

\begin{abstract}
Recently, several studies have investigated synchronization in quantum-mechanical limit-cycle oscillators. However, the quantum nature of these systems remained partially hidden, since the dynamics of the oscillator's phase was overdamped and therefore incoherent. We show that there exist regimes of underdamped and even quantum-coherent phase motion, opening up new possibilities to study quantum synchronization dynamics. To this end, we investigate the Van der Pol oscillator (a paradigm for a self-oscillating system) synchronized to an external drive. We derive an effective quantum model which fully describes the regime of underdamped phase motion and additionally allows us to identify the quality of quantum coherence. Finally, we identify quantum limit cycles of the phase itself.
\end{abstract}

\maketitle

{\bf\emph{Introduction.--}}
%
Synchronization is commonly studied in so-called limit cycle (LC) oscillators that arise from an interplay of linear and nonlinear effects \cite{2001_Kurths_Synchronization}. For instance, linear amplification causes an instability, whereas nonlinear damping limits the oscillator's dynamics to a finite amplitude. Notably, the phase remains free, which allows synchronization of the oscillator to an external periodic drive or other LC oscillators. A transition from the intrinsic LC motion towards synchronized oscillations occurs depending on the coupling strength to (and frequency mismatch of) the external reference. 

Quantum synchronization, i.e., the study of quantum systems whose classical counterparts synchronize, has recently attracted increasing theoretical attention. So far, studies of quantum synchronization have only explored overdamped phase motion. This implies that the dynamics, although taking place in quantum systems, remains always incoherent and classical-like, ruling out the observation of interesting effects like quantum tunnelling or superposition states of different synchronization phases. In the present article, we discover quantum-coherent phase dynamics.

Theoretical studies of quantum synchronization have been performed for different platforms, including optomechanics~\cite{2013_Ludwig,2016_Weiss_OMsync_noiseInducedEffects}, atoms and ions  \cite{2014_Minghui_QuantumSync_Atoms,2014_Hush_SpinCorr_QSync}, Van der Pol (VdP) oscillators~\cite{2013_Lee_QuantumSync_VdP_Ions, 2014_Walter_QuantumSync_1Vdp, 2014_Lee_EntanglementTongue_QSync, 2014_Walter_SynchronizationVdPosc,2016_Loerch_quantumSync_vdPKerr}, and superconducting devices~\cite{2008_Vinokur_Qsync,2013_Hriscu_QSync_SuperconductingDevice}. Measures of synchronization in the presence of quantum noise have been proposed in Refs.~\cite{2013_Mari_QuantumSync,2014_Hush_SpinCorr_QSync,2015_Ameri_MutualInformation_QuantumSync}. 

On the experimental side, only classical synchronization has been studied so far for a wide range of systems \cite{2005_Acebron_KuramotoReview}, including more recently  optomechanical systems \cite{2012_ZhangLipson_SynchronizationPRL,2013_BagheriTang_Synchronization, 2015_Shlomi_SyncExtDrive,2015_Lipson_ArraySync}. In the well-developed field of classical synchronization, overdamped phase motion is the standard ingredient both of phenomenological equations and microscopically derived models. For example, locking to an external force is described by the so-called Adler equation, a first-order differential equation for the phase. Similarly, synchronized optomechanical systems are described by the first-order phase equation of the Hopf-Kuramoto model~\cite{2011_Heinrich_CollectiveDynamics,2013_Ludwig,2014_Lauter_PhasePatterns}. However, it has been noticed that classical synchronization also allows for underdamped phase dynamics. For instance, the classical VdP oscillator features underdamped phase motion and even (synchronized) phase self-oscillations \cite{1988_Chakraborty_VdP_PhaseSync, 1990_Aronson_VdP_PhaseSync,2000_Pikovsky_PhaseSync, 2001_Kurths_Synchronization}. Both regimes have recently been observed experimentally using a nanoelectromechanical system~\cite{2014_Barois_SOSO}. Classical phase self-oscillations, also called phase trapping, have also been observed with coupled laser modes \cite{2011_Thevenin_PhaseTrapping_Laser}. Furthermore, synchronized Josephson junction arrays can be mapped to the Kuramoto model including inertia~\cite{2005_Trees_SyncJosephsonJunctions,2005_Acebron_KuramotoReview}. 

Here we will show that a regime of quantum-coherent dynamics exists and that underdamped phase dynamics is a necessary but not sufficient condition to observe this regime. Rather, it is the dynamically generated non-equilibrium dephasing rate that has to become smaller than the oscillation frequency. Additionally, we identify phase self-oscillations in the quantum regime. 

We obtain these insights for a paradigmatic model, the quantum version of the VdP oscillator subject to an external drive. The synchronization of the VdP oscillator to this external drive is an excellent test case to investigate universal synchronization behaviour. We derive an effective quantum model that captures the regime of underdamped phase dynamics. This allows us to identify a quality factor for the quantum coherence. We illustrate the potentially long coherence times by showing that initial negativities of a Wigner density vanish slowly. Finally, we briefly discuss possible experimental realizations.

{\bf\emph{Quantum model.--}}
%
\begin{figure}[ht]
\begin{center}
	\includegraphics[width=1\columnwidth]{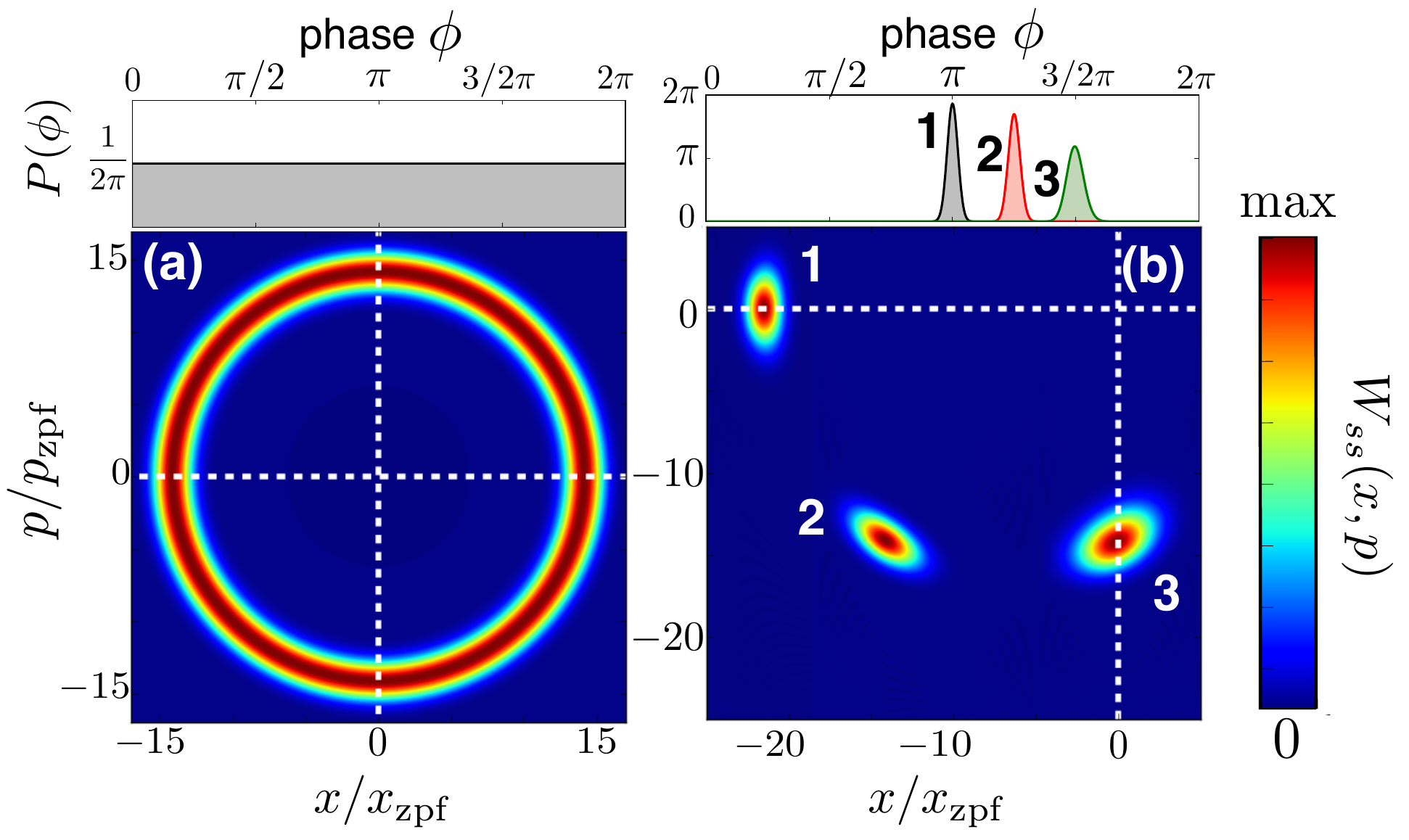}
	\caption{(color online).
	\emph{Quantum synchronization.}
	Steady-state Wigner density $W_{ss}(x,p)$ and phase probability distribution $P(\phi)$ of (a) an undriven ($F/\gamma_1=0$) and (b) an externally driven ($F/\gamma_1=10$) VdP oscillator. (a) The ring-shaped Wigner function indicates LC motion. (b) With increasing detuning $\Delta/\gamma_{1}$, the synchronization phase changes and synchronization becomes weaker. Parameters: $\gamma_2/\gamma_1=5\times10^{-3}$, (a) $\Delta/\gamma_1=0$, (b) ``1, 2, 3'' correspond to $\Delta/\gamma_1=0, 0.5, \text{ and }1$.}
	\label{fig_quRegimes}
\end{center}
\end{figure}
The quantum VdP oscillator subject to an external drive is described by the master equation ($\hbar = 1$)
\begin{align}\label{eq_MeqVdP}
	\dot{\hat{\rho}} = -i\left[-\Delta\hat{b}^{\dagger}\hat{b}+iF(\hat{b}-\hat{b}^{\dagger}), \hat{\rho}\right] + \gamma_{1} \mathcal{D}[\hat{b}^{\dag}]\hat{\rho} + \gamma_{2} \mathcal{D}[\hat{b}^{2}]\hat{\rho} \, ,
\end{align}
with $\mathcal{D}[\hat{O}]\hat{\rho}=\hat{O}\hat{\rho}\hat{O}^{\dagger} - \{ \hat{O}^{\dagger}\hat{O}, \hat{\rho} \}/2$. Here, $\Delta=\omega_{d}-\omega_0$ is the detuning of the oscillator's natural frequency $\omega_0$ from the frequency of the external drive $\omega_{d}$ and $F$ is the driving force. The two dissipative terms in Eq.~(\ref{eq_MeqVdP}) describe gain and loss of one and two quanta at rates $\gamma_{1}$ and $\gamma_{2}$, respectively.

In Fig.~\ref{fig_quRegimes} we show the steady-state Wigner function along with the corresponding phase probability distribution $P(\phi)=\sum_{n,m=0}^\infty \frac{e^{i(m-n)\phi}}{2\pi} \langle n|{\hat{\rho}}_{ss}|m\rangle$ \cite{2014_Hush_SpinCorr_QSync} by numerically solving Eq.~(\ref{eq_MeqVdP}) for its steady state ${\hat{\rho}}_{ss}$. In the absence of an external force ($F=0$), the two competing dissipation rates $\gamma_{1}$ and $\gamma_{2}$ lead to LC motion of the VdP oscillator, Fig.~\ref{fig_quRegimes}(a). For a finite applied force ($F\neq0$) and sufficiently small detuning $\Delta$, the VdP oscillator synchronizes to the external force and a fixed phase relation between the VdP oscillator and the force is present. In the rotating frame, this corresponds to a localized Wigner density and a phase distribution $P(\phi)$ with a distinct peak. With increasing detuning, the VdP oscillator is less synchronized to the external force [related to the height and width of $P(\phi)$] and the synchronization phase [peak position of $P(\phi)$] is shifted, Fig.~\ref{fig_quRegimes}(b).

These steady-state properties do not provide any information on the underlying synchronization dynamics, especially if we are trying to discover possible underdamped and quantum-coherent phase dynamics. To test for these regimes, we now derive an effective quantum model.

{\bf\emph{Effective quantum model.--}}
%
\begin{figure}[ht]
\begin{center}
	\includegraphics[width=0.99\columnwidth]{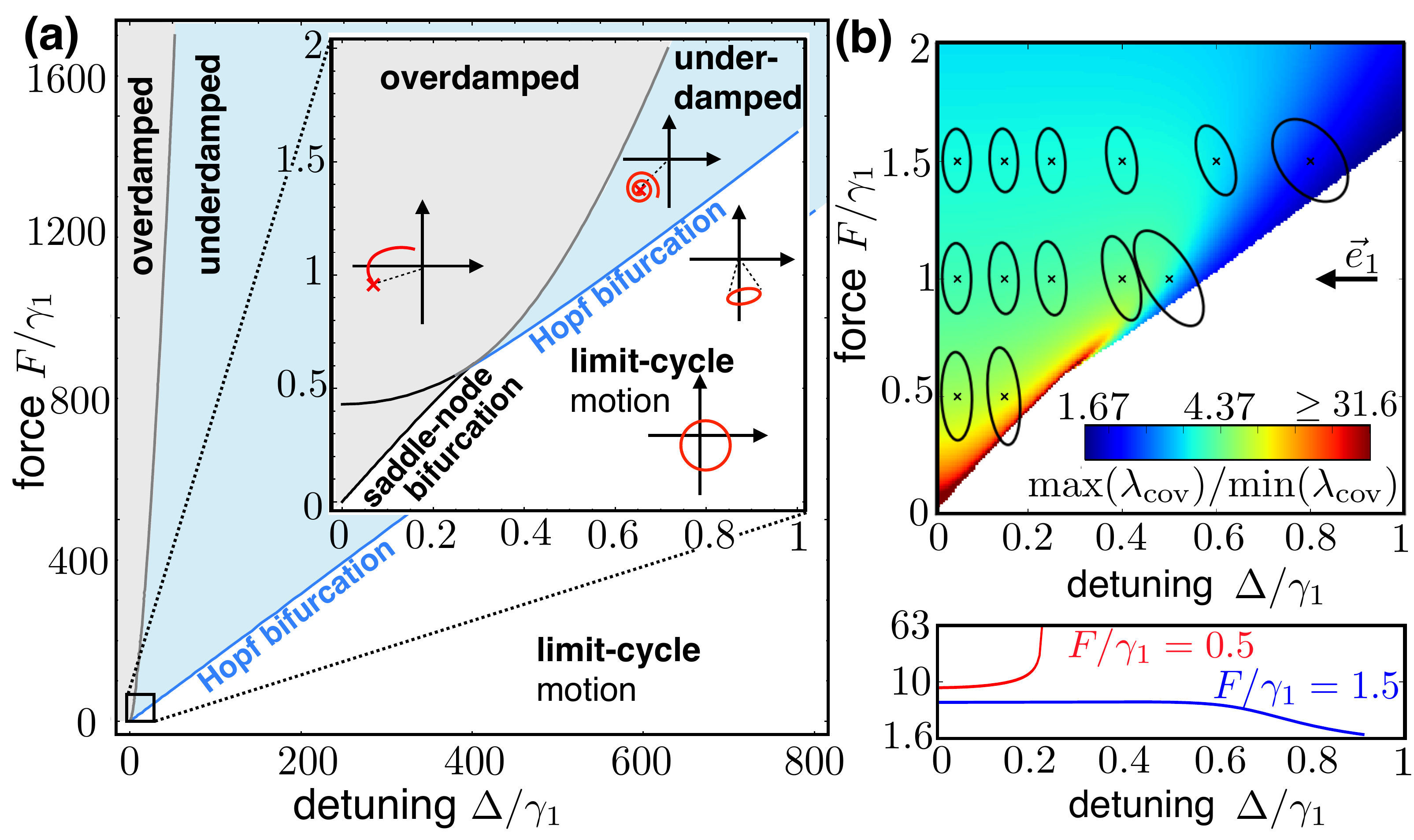}
	\caption{(color online).
	\emph{Classical phase diagram and squeezing.}
	(a) Overview of the classical synchronization regimes with sketches of typical phase-space trajectories.
	(b) Asymmetry of the steady-state squeezing ellipses, $\trm{max}(\lambda_\trm{cov})/\trm{min}(\lambda_\trm{cov})$, obtained from the effective model~\cite{footnote1}. At the black crosses we show the squeezing ellipses (not to scale) with their radial direction aligned along $\vec{e}_{1}$. Two cuts at different forcing are shown below the figure. Parameters: $\gamma_2/\gamma_1=0.1$.}
	\label{fig_squeezing}
\end{center}
\end{figure}
In the synchronized regime, the classical equation of motion for $\langle\hat{b}\rangle=\beta=Re^{i\phi}$,
\begin{align}\label{eq_vdPcomplexAmplitudeEq}
	\dot{\beta}=i\Delta\beta+\frac{\gamma_{1}}{2}\beta-\gamma_{2}|\beta|^{2}\beta-F \, ,
\end{align}
has a stable fixed point $\beta_{ss}=R_{ss}e^{i\phi_{ss}}$. We linearize Eq.~(\ref{eq_MeqVdP}) around $\beta_{ss}$ by defining $\hat{b} = \beta_{ss} + \delta\hat{b}$, where $\delta\hat b$ describes fluctuations around $\beta_{ss}$.
Neglecting terms of order $\mathcal{O}(\delta \hat{b}^{3})$ and higher, we obtain
\begin{align}\label{eq_eff_Meq}
	\dot{{\hat{\rho}}}_{\trm{eff}} = -i \left[ \hat{H}_{\trm{eff}}, {\hat{\rho}}_{\trm{eff}} \right] + \gamma_{1}\mathcal{D}[\delta\hat{b}^{\dagger}]{\hat{\rho}}_{\trm{eff}} + 4\gamma_{2}\left|\beta_{ss}\right|^{2}\mathcal{D}[\delta\hat{b}]{\hat{\rho}}_{\trm{eff}} \, ,
\end{align}
with the effective Hamiltonian
\begin{align}\label{eq_eff_1quantumVdP_Hamiltonian}
	\hat{H}_{\text{eff}}=-\Delta\delta\hat{b}^{\dagger}\delta\hat{b}-i\frac{\gamma_{2}}{2}\left(\beta_{ss}^{2}\delta\hat{b}^{\dagger}\delta\hat{b}^{\dagger}-\beta_{ss}^{*2}\delta\hat{b}\delta\hat{b}\right) \, .
\end{align}
This effective model captures the major features of the full quantum model and thus allows at least qualitative predictions about the behaviour of the system, while quantitative agreement varies with parameters. A comparison of the outcomes of Eqs.~(\ref{eq_MeqVdP}) and (\ref{eq_eff_Meq}) can be found in the Supplemental Material~\cite{supp2016}. The effective model is a squeezing Hamiltonian where the amount of squeezing depends on the classical steady-state solution $\beta_{ss}$. 
%
\begin{figure*}[t]
\begin{center}
	\includegraphics[width=0.99\textwidth]{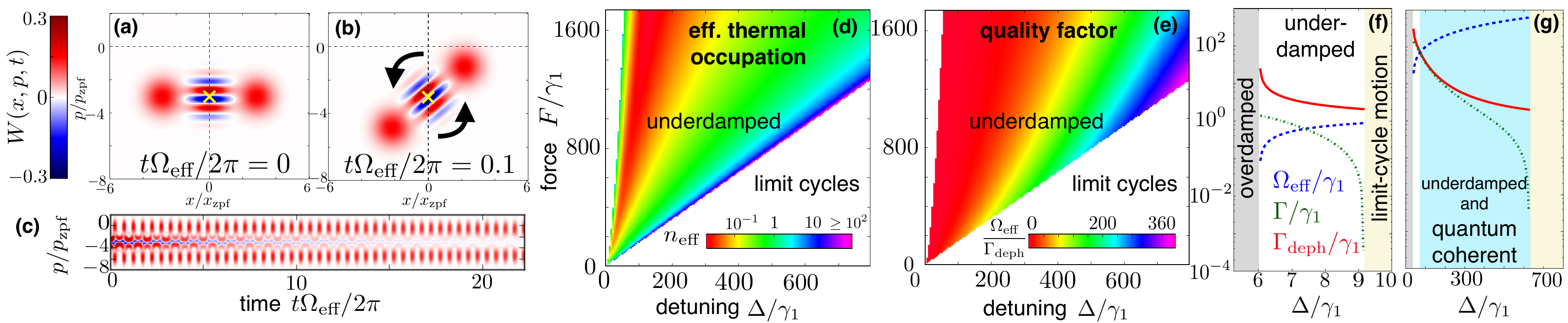}
	\caption{(color online).
	\emph{Quantum coherence.}
	Wigner densities $W(x,p,t)$ of (a) the initial state $\left|\Psi(t=0)\right\rangle \sim\left(\left|\beta_{ss}+2\right\rangle +\left|\beta_{ss}-2\right\rangle \right)$
	and (b) at a later time. The underdamped phase dynamics rotates the state around the classical steady-state solution (yellow cross).
	(c) Wigner density $W(p,x=0, t)$ with negativities that remain visible for many oscillations.
	(d) Effective temperature in the underdamped regime, indicated by $n_\trm{eff}$. The left white area corresponds to the overdamped regime.
	(e) Quality factor $\Omega_\trm{eff}/\Gamma_\trm{deph}$ in the underdamped regime. 
	(f) and (g) show the effective oscillation frequency $\Omega_\trm{eff}$ (dashed blue), damping $\Gamma$ (dash-dotted green), and dephasing rate $\Gamma_\trm{deph}$ (red) as a function of the detuning $\Delta$. 
	(f) At small force $F/\gamma_1=1.5$ the dephasing remains the dominant rate. 
	(g) In contrast, at larger force $F/\gamma_1=10^3$ the frequency $\Omega_\trm{eff}$ can significantly exceed both the dephasing and the damping (quantum-coherent regime).
	Parameters: $\gamma_2/\gamma_1=0.1$, and (a)-(c) $F/\gamma_1=1.5\times 10^3$ and $\Delta/\gamma_1=7\times 10^2$.
	(a)-(c) show numerical solutions to the full model Eq.~(\ref{eq_MeqVdP}), while (d)-(g) show the rates obtained from our effective quantum model which characterize the behaviour of the full system.
	}
	\label{fig_negWigner}
\end{center}
\end{figure*}
Diagonalizing Eq.~(\ref{eq_eff_1quantumVdP_Hamiltonian}) leads to 
\begin{align}
	\hat H_\trm{diag}=-\Omega_\trm{eff}\hat c^\dagger\hat c+\trm{const.} \, . 
\end{align}
Here, $\delta\hat b e^{-i\theta/2}=\trm{cosh}(\chi)\hat c+\trm{sinh}(\chi)\hat c^\dagger$, $A e^{i\theta}:=-i\gamma_2\beta_{ss}^2/2$, $\trm{tanh}(2\chi)=2 A/\Delta$, and  $\Omega_\trm{eff}=\sqrt{\Delta^2-4A^2}$ is the effective frequency. The corresponding master equation reads
\begin{align}
	\dot{{\hat{\rho}}}_{\trm{diag}} = -i \left[ \hat{H}_{\trm{diag}}, {\hat{\rho}}_{\trm{diag}} \right] + \Gamma_\uparrow\mathcal{D}[\hat{c}^{\dagger}]{\hat{\rho}}_{\trm{diag}} + \Gamma_\downarrow\mathcal{D}[\hat{c}]{\hat{\rho}}_{\trm{diag}} \, , 
\end{align}
with $\Gamma_\uparrow=4\gamma_2\left|\beta_{ss}\right|^2\trm{sinh}^2(\chi)+\gamma_1\trm{cosh}^2(\chi)$, $\Gamma_\downarrow=4\gamma_2\left|\beta_{ss}\right|^2\trm{cosh}^2(\chi)+\gamma_1\trm{sinh}^2(\chi)$, and neglecting fast rotating terms, such as $\hat c\hat c{\hat{\rho}}_\trm{eff}$. The diagonalized, effective model is a damped harmonic oscillator with frequency $\Omega_\trm{eff}$ and damping $\Gamma=\Gamma_\downarrow-\Gamma_\uparrow$. This unambiguously allows us to identify an underdamped phase dynamics regime following the standard procedure for a harmonic oscillator, i.e.,~we require $\Delta^2>4A^2$, which leads to a real-valued effective frequency $\Omega_\trm{eff}$. This is consistent with the corresponding classical dynamics derived from Eq.~(\ref{eq_eff_Meq}), leading to a {\emph{second}}-order differential equation of the phase,
\begin{align}
	\delta \ddot{\phi} + \Gamma \delta \dot{\phi} + \Omega^{2} \delta \phi = 0 \, . 
\end{align}
Here, $\delta\phi=\phi-\phi_{ss}$ describes phase deviations from the steady-state phase $\phi_{ss}$ and $\Omega=\sqrt{\Delta^{2}+\left(\gamma_{2}R_{ss}^{2}-\gamma_{1}/2\right)\left(3\gamma_{2}R_{ss}^{2}-\gamma_{1}/2\right)}$ is the bare frequency which is related to the effective frequency $\Omega_\trm{eff}=\sqrt{\Omega^2-\Gamma^2/4}=\sqrt{\Delta^2-\gamma_2^2|\beta_{ss}|^4}$; cf.~\cite{2014_Barois_SOSO, supp2016}.

Before we discuss results from our effective quantum model, we briefly review the relevant features of the corresponding classical ``phase diagram'' of synchronization, Fig.~\ref{fig_squeezing}(a). 
This phase diagram and its quantum analogue will help us to identify the parameter regime of underdamped phase motion, where we then can check for quantum coherence.
We obtain the boundaries between the regimes of the classical phase dynamics from a linear stability analysis of Eq.~(\ref{eq_vdPcomplexAmplitudeEq}); cf.~\cite{2001_Kurths_Synchronization,supp2016}. Notably, we distinguish two qualitatively different transitions from synchronization to no synchronization: 
At small forces, a saddle-node bifurcation characterizes the transition from synchronized (overdamped) dynamics directly to the LC regime. At larger forces, a regime of underdamped phase motion opens up before a Hopf bifurcation marks the onset of a LC which does not necessarily encircle the origin. 

Since we are actually interested in a quantum regime, it is worthwhile to see that these two qualitatively different transitions have also important consequences for the quantum dynamics.  In particular, we find a qualitative change of behaviour in the squeezing properties of the steady state. 
Since $\hat{H}_{\trm{eff}}$ is quadratic in $\delta\hat{b}$ the system is fully characterized by its covariance matrix $\sigma_{ij} = \trm{Tr}[{\hat{\rho}}_{\trm{eff}} \{ \hat{X}_{i}, \hat{X}_{j} \}/2]$ with the quadratures $\hat{X}_{1} = (\delta \hat{b} + \delta \hat{b}^{\dag})/\sqrt{2}$ and $\hat{X}_{2} = -i (\delta \hat{b} - \delta \hat{b}^{\dag})/\sqrt{2}$. The eigenvalues $\lambda_\trm{cov}$ of the covariance matrix determine the shape of the squeezing ellipse~\cite{2008_WallsMilburn_QuantumOptics}.
Their ratio (the asymmetry of the ellipses) is shown in Fig.~\ref{fig_squeezing}(b). Notably, at small forces, it increases with larger detuning. In contrast, at larger forces where we predict underdamped phase dynamics, the ellipses become more circular while increasing the detuning $\Delta$. 
Thus, the squeezing behaviour can be used as an indicator for the existence of a quantum regime of underdamped phase motion. 
The effective model becomes unstable if $\Gamma=0$, which corresponds to the classical fixed point losing its stability. Additional details on squeezing can be found in the Supplemental Material \cite{supp2016}.

{\bf\emph{Quantum coherence.--}}
Studying the effective model, we have identified the quantum regime of underdamped phase motion. Now we demonstrate that within this regime, it is possible to preserve quantum coherence for a significant time. To this end, we choose an initial state which possesses negativities in its Wigner function, Fig.~\ref{fig_negWigner}(a), and show that these negativities persist for a long time compared to the characteristic timescale of the dynamics $\Omega_\trm{eff}^{-1}$. The dynamics due to Eq.~(\ref{eq_MeqVdP}) leads to a rotation of the state around the classical steady state $\beta_{ss}$, Fig.~\ref{fig_negWigner}(b). Notably, this dynamical evolution has little influence on the coherence and the negativities of the Wigner density survive many oscillations of the system, Fig.~\ref{fig_negWigner}(c). After the loss of coherence, the state remains in a classical mixture of two displaced states and settles into the steady state only on an even longer timescale; see Ref.~\cite{supp2016} for a complete overview. All Wigner densities in Figs.~\ref{fig_negWigner}(a)-\ref{fig_negWigner}(c) are obtained by numerically solving the full master equation (\ref{eq_MeqVdP}).

This behaviour is successfully predicted by our effective model, which allows us to quantify quantum coherence within the underdamped regime and eventually identify a \emph{quantum-coherent regime}. The time scale on which the quantum system approaches the steady state is approximately given by the damping $\Gamma$. Thus, a \emph{necessary} condition to observe quantum-coherent motion is $\Omega_\trm{eff}>\Gamma$. Approaching the classical Hopf bifurcation,
the damping $\Gamma$ becomes arbitrarily small. However, a small damping does not imply a small dephasing rate $\Gamma_\trm{deph}=\Gamma_\uparrow+\Gamma_\downarrow$. The dephasing rate ultimately determines the lifetime of negativities, i.e., quantum coherence. With $\Gamma_\uparrow=\Gamma n_\trm{eff}$ and $\Gamma_\downarrow=\Gamma(n_\trm{eff}+1)$, the dephasing rate $\Gamma_\trm{deph}$ depends on both the damping $\Gamma$ and the effective occupation of the VdP oscillator $n_\trm{eff}$. This effective occupation comes about due to the driven-dissipative character of the quantum oscillator even at zero environmental temperature, also called quantum heating \cite{2011_Dykman_QuantumHeating}. It increases towards the boundaries of the underdamped regime, Fig.~\ref{fig_negWigner}(d), counteracting the decreasing damping. 
Additional insight is obtained by identifying a quality factor for quantum coherence, $\Omega_\trm{eff}/\Gamma_\trm{deph}$, which determines the lifetime of negativities in the Wigner density. Close to the instability and, more importantly, at large forcing and detuning, $\Omega_\trm{eff}/\Gamma_\trm{deph}$ increases and can become significantly larger than $1$, Fig.~\ref{fig_negWigner}(e). This is the quantum-coherent regime where negativities of the Wigner density can survive many oscillations of the system, Fig.~\ref{fig_negWigner}(c). 
Regarding Fig.~\ref{fig_negWigner}(e), the only remaining dimensionless parameter (apart from the normalized force and detuning) is the ratio of the damping rates $\gamma_2/\gamma_1$. It influences the region of stability of the effective model. For instance, increasing $\gamma_2/\gamma_1$ shifts the instability ($\Gamma=0$) to larger detuning. This allows to achieve a comparable quality factor $\Omega_\trm{eff}/\Gamma_\trm{deph}$ at smaller forcing but similar detuning - mainly because $\Omega_\trm{eff}$ increases with $\Delta$. %
In Figs.~\ref{fig_negWigner}(f) and (g) we show all relevant rates in the underdamped regime at small and large forcings, respectively. In both cases $\Omega_\trm{eff}$ increases, while $\Gamma$ and $\Gamma_\trm{deph}$ decrease with larger detuning. At small force, Fig.~\ref{fig_negWigner}(f), the dephasing rate remains the largest rate in the entire underdamped regime. Notably, for large $F$, Fig.~\ref{fig_negWigner}(g), we find that $\Omega_\trm{eff}$ can become significantly larger than both $\Gamma$ and $\Gamma_\trm{deph}$, thus entering the quantum-coherent regime. This is the key element to observing long-lived quantum coherence.

\begin{figure}[t]
\begin{center}
	\includegraphics[width=0.99\columnwidth]{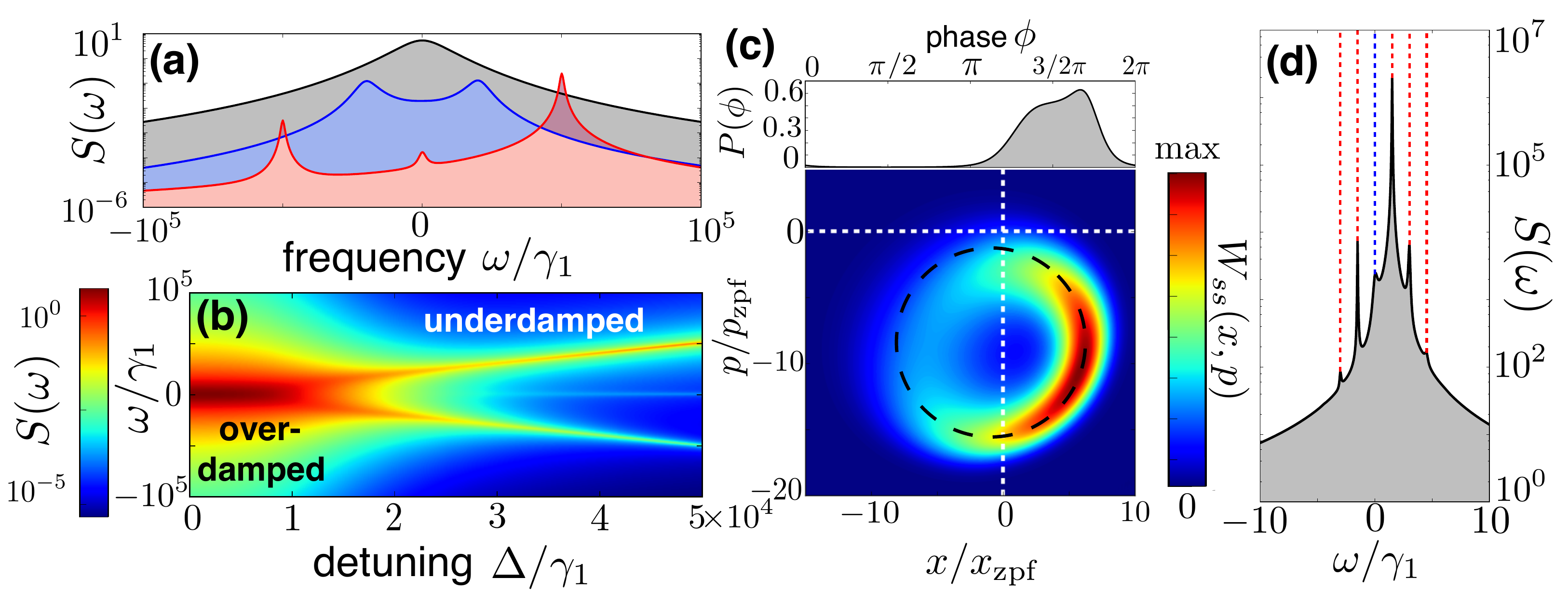}
	\caption{(color online).
	\emph{Spectrum.}
	(a) and (b) show the spectrum $S(\omega)$ of a synchronized VdP oscillator for different detunings $\Delta/\gamma_1=0$ (upper black spectrum), $\Delta/\gamma_1=2\times10^4$ (middle blue spectrum), and $\Delta/\gamma_1=5\times10^4$ (lower red spectrum). In the overdamped regime (upper black) the spectrum shows a single peak at $\omega=0$, while in the underdamped regime (middle blue and lower red) double peaks at $\pm\Omega_\trm{eff}$ emerge. Parameters: $\gamma_2/\gamma_1=2\times 10^3, F/\gamma_1=2\times 10^4$. (c) Steady-state Wigner density $W_{ss}(x,p)$ and phase probability distribution $P(\phi)$ of a VdP oscillator showing phase self-oscillations, i.e., a ring-like Wigner density not encircling the origin (indicated by the dashed black line). The corresponding spectrum (d) shows multiple peaks at higher harmonics. Parameters for (c) and (d): $\gamma_2/\gamma_1=5\times10^{-3}, F/\gamma_1=10, \Delta/\gamma_1=1.55$. All subfigures (a)-(d) show numerical solutions of the full model Eq.~(\ref{eq_MeqVdP}).
	}
	\label{fig_spectrum}
\end{center}
\end{figure}

{\bf\emph{Spectrum.--}}
To shed more light on the possibility to experimentally observe the transition from overdamped to underdamped synchronization dynamics, we investigate the spectrum $S(\omega)=\int_{-\infty}^\infty dt e^{i\omega t}\langle\hat b^\dagger(t)\hat b(0)\rangle$. We obtain $S(\omega)$ from the steady state of Eq.~(\ref{eq_MeqVdP}) by applying the quantum regression theorem or analytically from the effective model; see Supplemental Material~\cite{supp2016}. The spectrum carries information on the frequencies of the driven VdP oscillator.
Figure~\ref{fig_spectrum}(a) shows $S(\omega)$ for a fixed external force and various detunings, corresponding to the overdamped (upper black spectrum) and underdamped (middle blue, lower red spectra) regime. In the overdamped regime the spectrum shows a single peak close to $\omega=0$, indicating synchronization to the external force. With increasing detuning, the spectrum develops from a single peak to two peaks which now sit at approximately $\pm\Omega_\trm{eff}$. A small remainder of the central peak at $\omega=0$ becomes visible for a larger splitting of the main peaks. The emerging double peaks clearly indicate the transition from overdamped to underdamped phase dynamics, Fig.~\ref{fig_spectrum}(b). The increasing asymmetry of $S(\omega)$ results from the coupling of amplitude and phase dynamics.
%
For even larger detuning, synchronization is lost which ultimately leads to a single peak in the spectrum at $\omega = \Delta$. 
A recent experiment synchronized two nanomechanical oscillators by coupling to a common cavity mode \cite{2013_BagheriTang_Synchronization}. Curiously, the cavity output spectrum showed sidebands next to the common frequency of the locked oscillators. These sidebands were suggested to arise from (classical) underdamped phase motion of the oscillators, which is also consistent with the classical limit of our theory.

Interestingly, we find that the phase can even undergo self-oscillations. In the quantum regime, these \emph{phase self-oscillations} appear (in analogy to the classical scenario) at the boundary of underdamped phase motion just before the loss of synchronization occurs. A circular LC opens up around the former stable fixed point. In the quantum regime this is smeared by quantum fluctuations and becomes visible only once the LC is large enough. If that LC expands even further, it will eventually come to resemble the original unsynchronized state: The LC encircles the origin of phase space and the corresponding phase distribution is flat, Fig.~\ref{fig_quRegimes}(a). However, in Fig.~\ref{fig_spectrum}(c), this is not yet the case, i.e., the LC does not encircle the origin. The oscillator has still a tendency to be locked to the phase of the external force. This is also reflected in the corresponding phase distribution $P(\phi)$ which becomes asymmetric and shows the onset of a double peak structure.  
Notably, \emph{phase self-oscillations} are accompanied by the appearance of a series of peaks in the spectrum, Fig.~\ref{fig_spectrum}(d), representing higher harmonics of the main phase-oscillation frequency.

{\bf\emph{Experimental realization.--}}
The regime of underdamped quantum phase motion and even quantum phase self-oscillations could be experimentally studied in a variety of systems.
For instance, trapped ions are promising candidate systems for studying synchronization in the quantum regime~\cite{2013_Lee_QuantumSync_VdP_Ions,2014_Lee_EntanglementTongue_QSync}. The possibility to prepare nonclassical states experimentally~\cite{1996_Leibfried_QuantumStatesTrappedIons} allows for probing the quantum-coherent nature of the underdamped phase dynamics. Based on parameters for trapped $^{171}\trm{Yb}^+$ ions \cite{2013_Lee_QuantumSync_VdP_Ions, 2012_Islam_PhDThesis, 2003_Leibfried_RevMod_Ions}, we estimate that it should be possible to observe significant quantum coherence. 
In this scenario, the negative and nonlinear damping are both of the order of  kHz, with $\gamma_2/\gamma_1\sim1$. To observe quantum-coherent underdamepd phase dynamics the detuning $\Delta$ and the external force $F$ should be a few hundred kHz each. This is realistic, with frequencies of the motional state in the MHz regime.  
Furthermore, mechanical self-oscillations in cavity optomechanics have been discussed theoretically~\cite{2006_FM_DynamicalMultistability} and observed experimentally~\cite{2005_KippenbergVahala_SelfOsc,2008_Metzger_SelfOsc}. Thus, they are well-suited to study synchronization, and classical synchronization phenomena have already been demonstrated experimentally~\cite{2012_ZhangLipson_SynchronizationPRL,2013_BagheriTang_Synchronization, 2015_Shlomi_SyncExtDrive,2015_Lipson_ArraySync}.
Yet another possible platform to observe quantum-coherent phase motion are superconducting 
microwave circuits. These are exceptional and highly tuneable platforms for experimentally investigating quantum systems. In principle, arbitrary quantum states can be realized~\cite{2008_Hofheinz_GeneratingFockStates,2008_Deleglise_EngineeringQuantumStates,2009_Hofheinz_ArbitraryQuantumStates}. Even, the faithful engineering of two-photon losses in such systems has been demonstrated~\cite{2015_Leghtas_EngineeredTwoPhotonLoss}. This makes them very interesting for studying quantum-coherent phase motion and phase self-oscillations of a quantum VdP oscillator.

{\bf\emph{Conclusion.--}}
We have shown that the phase of a synchronized quantum Van der Pol oscillator exhibits intriguing underdamped and even quantum-coherent phase
dynamics around the synchronized steady state. In order to explore this interesting regime, we have developed an effective quantum model and identified where the dephasing rate becomes sufficiently small to observe quantum-coherent phase motion. As a direct consequence, we have shown that this preserves a nonclassical quantum state for many phase oscillations. We estimate that this could readily be observed in state-of-the-art experiments. 
While we have analyzed the simplest synchronization phenomenon, to an external drive, the regime identified here will also show up in the quantum phase dynamics of two coupled oscillators or even lattices \cite{2013_Ludwig}. In the latter case, phenomena such as quantum motion of phase vortices may potentially become observable.

\begin{acknowledgements}
\emph{Acknowledgements:} We acknowledge financial support by the Marie Curie ITN cQOM and the ERC OPTOMECH.
\end{acknowledgements}

\bibliographystyle{apsrev4-1}

\clearpage


\onecolumngrid
\setcounter{figure}{0}
\setcounter{equation}{0}
\renewcommand{\thefigure}{S\arabic{figure}}
\renewcommand{\theequation}{S\arabic{equation}}

\begin{center}
        \large{\bf{Supplemental Material for ``Quantum-coherent phase oscillations in synchronization''}}
\end{center}
\begin{center}
        Talitha~Weiss, Stefan~Walter, and Florian~Marquardt \\
        \emph{\small{Friedrich-Alexander University Erlangen-N{\"u}rnberg (FAU), Department of Physics, Staudtstra{\ss}e 7, 91058 Erlangen, Germany}}\\
        \emph{\small{and Max Planck Institute for the Science of Light,
Staudtstra{\ss}e 2, 91058 Erlangen, Germany}}
\end{center}

%
\section{Details on the classical synchronization phase diagram}
%
\begin{figure}[b]
\begin{center}
	\includegraphics[width=0.99\columnwidth]{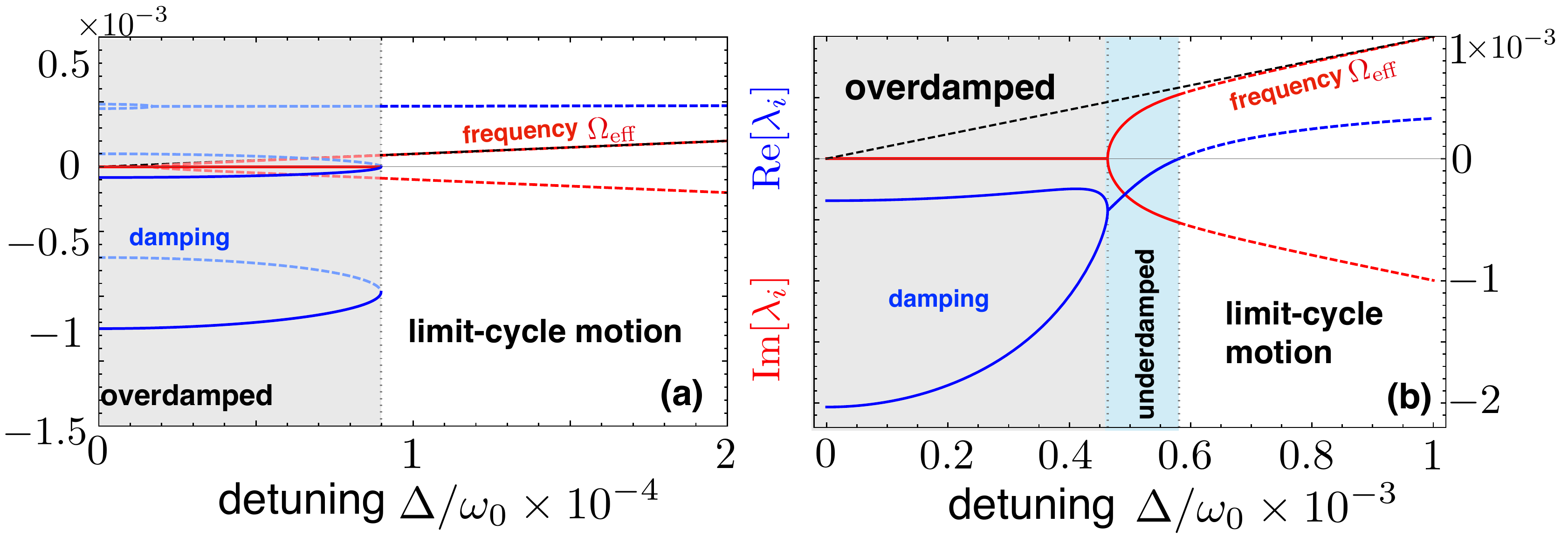}
	\caption{(color online).
	\emph{Classical synchronization transitions.}
	 (a) and (b) show the real (blue) and imaginary (red) parts of the eigenvalues $\lambda_i$ from the linear stability analysis that are related to the damping
	 and the frequency respectively. The dashed blue and red lines correspond to unstable fixed points, while the solid lines correspond to a stable fixed point. Each fixed point has two associated eigenvalues that might be degenerate.The black dashed line indicates the frequency of the free VdP oscillator in the rotating frame, $\Delta$. 
	(a) At small forcing, $F/\gamma_{1}=0.2$, the VdP oscillator starts out in a regime of synchronization and overdamped phase dynamics and transitions to limit-cycle motion via a saddle-node bifurcation when increasing the detuning $\Delta$. 
	(b) At larger external force, $F/\gamma_{1}=1$, the transition from synchronization to no synchronization occurs via a Hopf bifurcation,
	 thus first crossing a region of underdamped phase motion.
	Parameters as in Fig.~2 of the main text. 
	}
	\label{fig_classicalBifurcations}
\end{center}
\end{figure}

Here, we discuss some additional details of the classical phase diagram of a Van der Pol (VdP) oscillator synchronizing to an external force, cf.~Fig.~2(a) of the main text.
The boundaries for the regimes of overdamped, underdamped, and limit-cycle motion are obtained from a linear stability analysis of Eq.~(2) of the main text (see for instance also Ref.~\cite{2001_Kurths_Synchronization} for more details). Using $\beta=\beta_{ss}+\delta\beta$ and keeping only first order terms of $\delta\beta$, the linearized equation of motion is 
\begin{align}\label{eq_linearizedDeltaBeta}
	\delta\dot\beta=i\Delta\delta\beta-\gamma_{2}\beta_{ss}^{2}\delta\beta^{*}+\frac{\gamma_{1}}{2}\delta\beta-2\gamma_{2}\left|\beta_{ss}\right|^{2}\delta\beta .
\end{align}
The eigenvalues $\lambda_i$ of the corresponding Jacobi matrix are related to damping and effective frequency of the VdP oscillator and contain information about the properties of the corresponding fixed point $\beta_{ss}$. 
In Fig.~\ref{fig_classicalBifurcations}, we show the real and imaginary part of the eigenvalues $\lambda_{1/2}=-2|\beta_{ss}|^2\gamma_2+\gamma_1/2\pm\sqrt{|\beta_{ss}|^4\gamma_2^2-\Delta^2}$. Note that, depending on parameters, Eq.~(\ref{eq_linearizedDeltaBeta}) features either one or three fixed points. However, if there are three fixed points, see Fig.~\ref{fig_classicalBifurcations}(a), only one of them is stable. We show the real and imaginary parts of the eigenvalues of a stable fixed point as solid lines and the eigenvalues corresponding to unstable fixed points as dashed lines. Note that in the synchronized regimes (both overdamped and underdamped) synchronization towards the stable fixed point occurs. 

The phase diagram, cf.~Fig.~2(a) of the main text, shows that the regime of limit-cycle motion can be entered via two different types of bifurcations which we show here in detail:
In Fig.~\ref{fig_classicalBifurcations}(a) the regime of limit-cycle motion is entered via a saddle-node bifurcation, while in Fig.~\ref{fig_classicalBifurcations}(b) a Hopf bifurcation marks the onset of limit-cycle oscillations. In both cases a single, unstable fixed point exists in the limit-cycle regime, i.e., the corresponding real parts of the eigenvalues, associated to damping, are positive. This indicates amplification and the stabilizing, nonlinear effects that lead to a limit cycle are not included in the linear stability analysis. 
At smaller detuning, before the limit-cycle regime, synchronization towards a stable fixed point occurs. 
The important difference between the two possible transitions is determined by the imaginary part of the eigenvalues of this stable fixed point. The imaginary part is related to the oscillation frequency and approaching the saddle-node bifurcation, the imaginary parts remain zero. This implies that, for detunings below the bifurcation, no characteristic oscillation frequency exists. The non-zero imaginary parts visible in Fig.~\ref{fig_classicalBifurcations}(a) belong to unstable fixed points, which are not of importance in the context of synchronization.
In contrast, at larger force, Fig.~\ref{fig_classicalBifurcations}(b), a Belyakov-Devaney transition \cite{2010_Broer_HandbookDynSys} occurs even before the Hopf bifurcation. There the real parts of the eigenvalues become equal but remain negative (i.e.~there exists a stable fixed point), yet the imaginary parts become nonzero. This implies that in this case the steady state is approached in an oscillatory fashion and determines the regime of underdamped phase motion. Only at even larger detuning a Hopf bifurcation occurs where the real parts become positive as well and a limit cycle is created. 

The boundaries between the different regimes of classical synchronization phase dynamics are determined by the following explicit expressions:
$(i)$ a saddle-node bifurcation given by $F^{2}=\frac{\left(\mp2\gamma_{1}+\sqrt{\gamma_{1}^{2}-12\Delta^{2}}\right)\left(-\gamma_{1}\sqrt{\gamma_{1}^{2}-12\Delta^{2}}\pm\left(12\Delta^{2}+\gamma_{1}^{2}\right)\right)}{108\gamma_{2}^{2}}$,
$(ii)$ a transition from a stable node to a stable focus (Belyakov-Devaney transition) which is defined by $F^{2}=\left(-\frac{\gamma_{1}}{2}+\Delta\right)^{2}\frac{\Delta}{\gamma_{2}}+\frac{\Delta^{3}}{\gamma_{2}}$
with $|\Delta|>\gamma_{1}/4$,
and $(iii)$ a Hopf bifurcation described by $F^{2}=\frac{1}{4}\frac{\gamma_{1}}{\gamma_{2}}\Delta^{2}+\frac{1}{64}\frac{\gamma_{1}^{3}}{\gamma_{2}}$
with $|\Delta|>\gamma_{1}/4$.
Note that this linear analysis does not allow us to distinguish between stable self-oscillations of the phase (limit cycles not evolving around the origin) and ordinary limit cycles where the phase is monotonously increasing.  

\section{Classical dynamics of the effective quantum model}

The effective quantum model, Eq.~(3) of the main text, allows us to discuss the corresponding classical dynamics which is given by
$\delta\dot{\beta}=\trm{Tr}[\delta\hat{b} \dot{\rho}_{\trm{eff}}] =i\Delta\delta\beta-\gamma_{2}\beta_{ss}^{2}\delta\beta^{*}+\frac{\gamma_{1}}{2}\delta\beta-2\gamma_{2}\left|\beta_{ss}\right|^{2}\delta\beta$. This is equivalent to the linearized equation (\ref{eq_linearizedDeltaBeta}) confirming that we have indeed derived the correct linearized \emph{quantum} model.
It is instructive to consider to first split the complex amplitude $\beta$ into amplitude $R$ and phase $\phi$ such that $\beta=R e^{i\phi}$, and then obtain the corresponding equations for the amplitude and phase deviations $\delta R$ and $\delta \phi$. These deviations are simply defined as the difference between the actual amplitude $R$ (phase $\phi$) from the steady-state amplitude $R_{ss}$ (steady-state phase $\phi_{ss}$), i.e.~$\delta R=R-R_{ss}$
and $\delta\phi=\phi-\phi_{ss}$.
Since $\delta R$ and $\delta \phi$ are small, $\delta R$ is approximately the change in direction of $R_{ss}$ and $\delta \phi$ is
approximately the change perpendicular to this. For $\delta \beta = re^{i \varphi}$ we then obtain $\delta \phi \approx r \sin(\varphi-\phi_{ss})$
and $\delta R \approx r \cos(\varphi-\phi_{ss})$ and with this
\begin{align}
	\delta \dot{\phi}	&=  \Delta\delta R-\left(\gamma_{2}R_{ss}^{2}-\frac{\gamma_{1}}{2}\right) \delta \phi \, , \label{eq_clFirstOrderDeq}\\
	\delta \dot{R}	&= -\left(3\gamma_{2}R_{ss}^{2}-\frac{\gamma_{1}}{2}\right) \delta R - \Delta \delta \phi \, , \label{eq_clFirstOrderDeq2}
\end{align}
which can be combined to a second-order differential equation for the phase, 
\begin{align}\label{eq_2ndorderPhase}
	\delta \ddot{\phi} + \Gamma \delta \dot{\phi} + \Omega^{2} \delta \phi = 0 \, .
\end{align}
Here we have defined $\Gamma=\left(4\gamma_{2}R_{ss}^{2}-\gamma_{1}\right)$
and $\Omega=\sqrt{\Delta^{2}+\left(\gamma_{2}R_{ss}^{2}-\frac{\gamma_{1}}{2}\right)\left(3\gamma_{2}R_{ss}^{2}-\frac{\gamma_{1}}{2}\right)}$. 
Notably Eq.~(\ref{eq_2ndorderPhase}) is describes a common harmonic oscillator which allows for overdamped as well as underdamped motion. The transition from overdamped to underdamped solutions is characterized
by $\Omega^{2}=\Gamma^{2}/4$, i.e.~where the effective oscillation frequency of the system $\Omega_\trm{eff}=\sqrt{\Omega^2-\Gamma^2/4}$ becomes real-valued. 
The solution to Eq.~(\ref{eq_2ndorderPhase}) becomes unstable if $\Gamma<0$, revealing the onset of limit-cycle motion. The limit-cycle motion itself depends on nonlinear effects to stabilize and thus cannot be described with the linearized equations.

The parameters $\Gamma$ and $\Omega_\trm{eff}$ obtained from this classical analysis are equal to the damping and effective frequency appearing in the effective quantum model.

\section{Comparison of the full and the effective quantum model}
\begin{figure}[t]
\begin{center}
	\includegraphics[width=0.5\columnwidth]{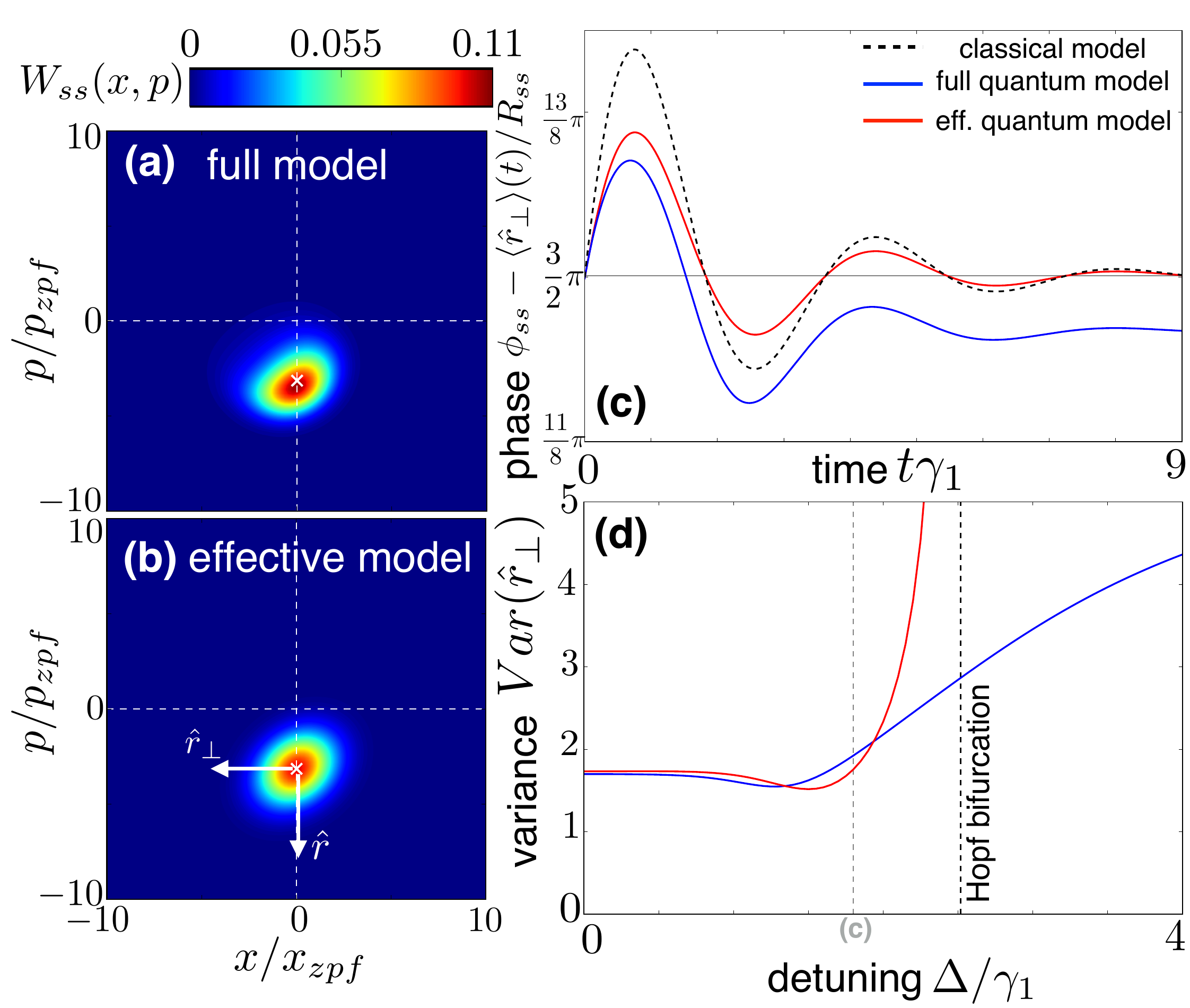}
	\caption{(color online).
	\emph{Full and effective quantum model.}
	Steady-state Wigner densities of (a) the full and (b) of the effective quantum model. The corresponding ``phase trajectories'' in (c) show similar oscillating behaviour, although relaxing to a different steady state. The black, dashed line gives the classical trajectory for comparison. In (d) we show the variance $Var(\hat{r}_\perp)$ as a function of the detuning. The deviations of the effective from the full model increase towards the Hopf bifurcation, where the effective model breaks down. Parameters: $\gamma_2/\gamma_1=0.1$, $F/\gamma_1=4$, and $\Delta/\gamma_1=1.8$. 
	}
	\label{fig_compareModels}
\end{center}
\end{figure}

Here we compare results from the full quantum model, Eq.~(1) of the main text, to results from the e§ffective model, Eq.~(3) of the main text, and the outcome of the classical equations (\ref{eq_clFirstOrderDeq}) and (\ref{eq_clFirstOrderDeq2}). In Fig.~\ref{fig_compareModels}(a) and (b) we show the steady-state Wigner density obtained from the full quantum model and the effective model respectively. The result of the effective model needs to be displaced to the classical steady state $\beta_{ss}$, indicated by the white cross. The Wigner densities obtained from the full and the effective quantum model match reasonably well. The parameters were chosen such that first deviations become visible: (i) The Wigner density of the full model is no longer centered exactly around the classical solution $\beta_{ss}$, while the effective model does so by construction. (ii) The effective quantum model is described by a squeezing Hamiltonian, cf.~main text. Thus the corresponding Wigner densities are ellipses, while the full model can lead to additional curvature in the Wigner density (more banana-shaped).

Within the effective model synchronization attracts the system's dynamics towards the stable fixed point $\beta_{ss}$. We can capture the dynamics using small deviations around $\beta_{ss}$. A natural choice are deviations in radial direction, $\delta R$, and in phase direction, $\delta \phi$, similar to the classical treatment. We can define corresponding operators $\hat r =\trm{cos}(\phi_{ss})\hat{x}/x_\trm{zpf}+\trm{sin}(\phi_{ss})\hat{p}/p_\trm{zpf}$ in radial direction and perpendicular to it, $\hat r_\perp= \trm{sin}(\phi_{ss})\hat{x}/x_\trm{zpf}-\trm{cos}(\phi_{ss})\hat{p}/p_\trm{zpf}$.
With this, deviations of the phase can be approximated via $\delta\phi\approx-\langle\hat r_\perp\rangle/R_{ss}$ such that the full phase is given by $\phi(t)\approx\phi_{ss}-\langle\hat r_\perp\rangle(t)/R_{ss}$.
We show the phase as a function of time in Fig.~\ref{fig_compareModels}(c). The system shows underdamped phase motion, i.e.~a few damped oscillations can be observed in the full and effective quantum model, as well as in the classical simulation. It is consistent with the corresponding Wigner densities, that the trajectories of the full and effective quantum model are damped towards a different steady state. Only the steady state of the effective quantum model and the classical equations are equal by construction. Note that the relation of $\langle\hat r_\perp\rangle$ to the phase deviations $\delta\phi$ is only accurate if the deviations are small. In Fig.~\ref{fig_compareModels}(d) we show the variance $Var(\hat r_\perp)$ as a function of detuning. For small $\Delta$ synchronization works best, i.e.~the Wigner density is more confined in phase space and thus the resulting variance is small. Deviations between the full and effective quantum model appear with increasing detuning. Then, synchronization becomes weaker and the full model can develop a less ellipse-like Wigner density. Approaching the Hopf bifurcation the variance within the effective model blows up, signalling the break-down of the model. The full model shows an increasing variance, which is consistent with the synchronization becoming weaker and the Wigner density becoming more smeared out. 

\section{Details on the squeezing}
\begin{figure}[b]
\begin{center}
	\includegraphics[width=0.5\columnwidth]{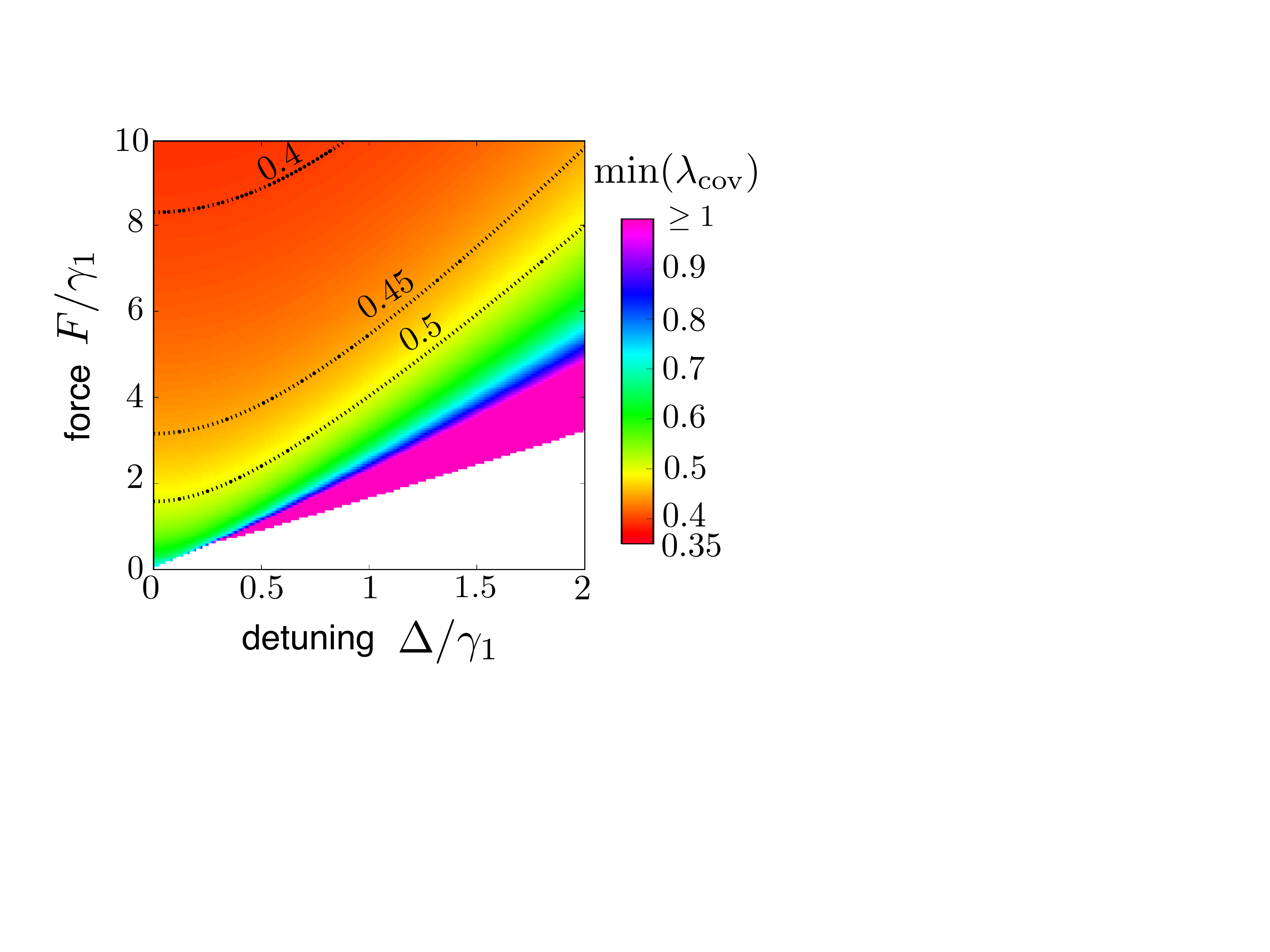}
	\caption{(color online).
	\emph{Squeezing.}
	Minimum of the eigenvalues $\lambda_\mathrm{cov}$ of the covariance matrix $\sigma$, as a function of the detuning $\Delta$ and forcing $F$. Values smaller than $0.5$ indicate squeezing below shot noise. Note that squeezing is strongest at small detuning and large forcing. Parameters: $\gamma_2/\gamma_1=0.1$.
	}
	\label{fig_belowShotNoise}
\end{center}
\end{figure}

In the main text we derived the squeezing Hamiltonian of our effective model, Eq.~(4) of the main text, and discussed the asymmetry of the squeezing ellipses in Fig.~2(b) of the main text. Due to the quadratic Hamiltonian, the state is fully characterized by its covariance matrix $\sigma_{ij} = \trm{Tr}[{\hat{\rho}}_{\trm{eff}} \{ \hat{X}_{i}, \hat{X}_{j} \}/2]$ with the quadratures $\hat{X}_{1} = (\delta \hat{b} + \delta \hat{b}^{\dag})/\sqrt{2}$ and $\hat{X}_{2} = -i (\delta \hat{b} - \delta \hat{b}^{\dag})/\sqrt{2}$ (same definition as in the main text). 
The equation of motion for the covariance matrix can be expressed in the following form,
\begin{equation}
	\dot\sigma=M\sigma + \sigma M^\top+D,
\end{equation}
with the matrices 
\begin{align}
	M&=\begin{pmatrix} i(r-r^*)+\gamma_1/2-2\gamma_2|\beta_{ss}|^2 & r+r^*-\Delta \\ r+r^*+\Delta & -i(r-r^*)+\gamma_1/2-2\gamma_2 |\beta_{ss}|^2\end{pmatrix}\\
	D&=\begin{pmatrix} \gamma_1/2+2\gamma_2|\beta_{ss}|^2&  \\ & \gamma_1/2+2\gamma_2|\beta_{ss}|^2 \end{pmatrix}.
\end{align} 
Here we used $r=i\gamma_2\beta_{ss}^2/2$ for brevity. The steady-state solution to this equation of motion, $\dot\sigma=0$, can be analytically obtained, resulting in a $2\times 2$ matrix $\sigma$ that depends only on system parameters and the classical steady-state amplitude $\beta_{ss}$. Then, the eigenvalues of the covariance matrix $\sigma$ can be calculated and analyzed.
The ratio of these eigenvalues determines the asymmetry of the squeezing ellipses discussed in the main text. However, also the \emph{absolute} amount of squeezing can be analyzed by comparing to the size of the vacuum state. If any direction of the squeezing ellipse becomes smaller than the width of the vacuum state this is referred to as squeezing below shot noise. To this end, we calculate the shot-noise covariance matrix $\sigma_{ij}^\trm{sn}=\trm{Tr}[\mid 0\rangle\langle 0\mid \{ \hat{X}_{i}, \hat{X}_{j} \}/2]$, which is diagonal and has $\lambda_\trm{sn}=1/2$ as doubly degenerate eigenvalue. We compare this to the smallest eigenvalue $\mathrm{min}(\lambda_\mathrm{cov})$ of the covariance matrix of the synchronized VdP oscillator. Squeezing below shot noise occurs for values $\trm{min}(\lambda_\trm{cov})<1/2$.

Interestingly, the synchronized VdP oscillator does feature squeezing below shot noise at small detuning and sufficiently large forcing.  As shown in Fig.~\ref{fig_belowShotNoise}, for a fixed detuning, squeezing becomes stronger if the external force $F$ is increased, eventually dropping below the shot noise value. In combination with Fig.~2(b) of the main text we conclude that approximately the radial direction is squeezed. Since squeezing below the shot noise level occurs mainly at small detuning, it occurs mostly within the regime of overdamped phase motion, but can reach into the underdamped regime as well. However, approaching the classical Hopf bifurcation, i.e., the instability of the effective model, by increasing the detuning $\Delta$, the squeezing necessarily decreases since the Wigner density smoothly transforms back into a (circular) limit cycle.

This parameter dependence of the absolute squeezing can also be directly explained from the squeezing Hamiltonian, Eq.~(4) of the main text. At first sight squeezing depends on the steady state of the system, i.e., $\beta_{ss}$, and thus has an intricate dependence on all parameters. However, we generally observe that large forcing $F$ leads to large values of $|\beta_{ss}|$. Notably, the squeezing Hamiltonian does not depend on this absolute value, but on the complex value $\beta_{ss}^2$ instead. Investigating the steady state Wigner density of the synchronized VdP oscillator, we observed in Fig.~1(b) of the main text a crucial dependence on the detuning: Although the value $|\beta_{ss}|$ does slightly decrease with $\Delta$, the more important effect is a rotation in phase space, corresponding to a change of the synchronization phase. Thereby $\beta_{ss}$ transforms from an almost real quantity to an almost purely imaginary quantity, thus significantly decreasing the real part of $\beta_{ss}^2$ even if its modulus would be conserved completely. 
Therefore, we can conclude that large squeezing appears if the force is sufficiently large compared to the detuning such that the synchronization phase (the peak position of the phase distribution) is close to the ideal value of $0$ or $\pi$.

\section{Long time evolution of coherent synchronization dynamics}

\begin{figure}[t]
\begin{center}
	\includegraphics[width=\columnwidth]{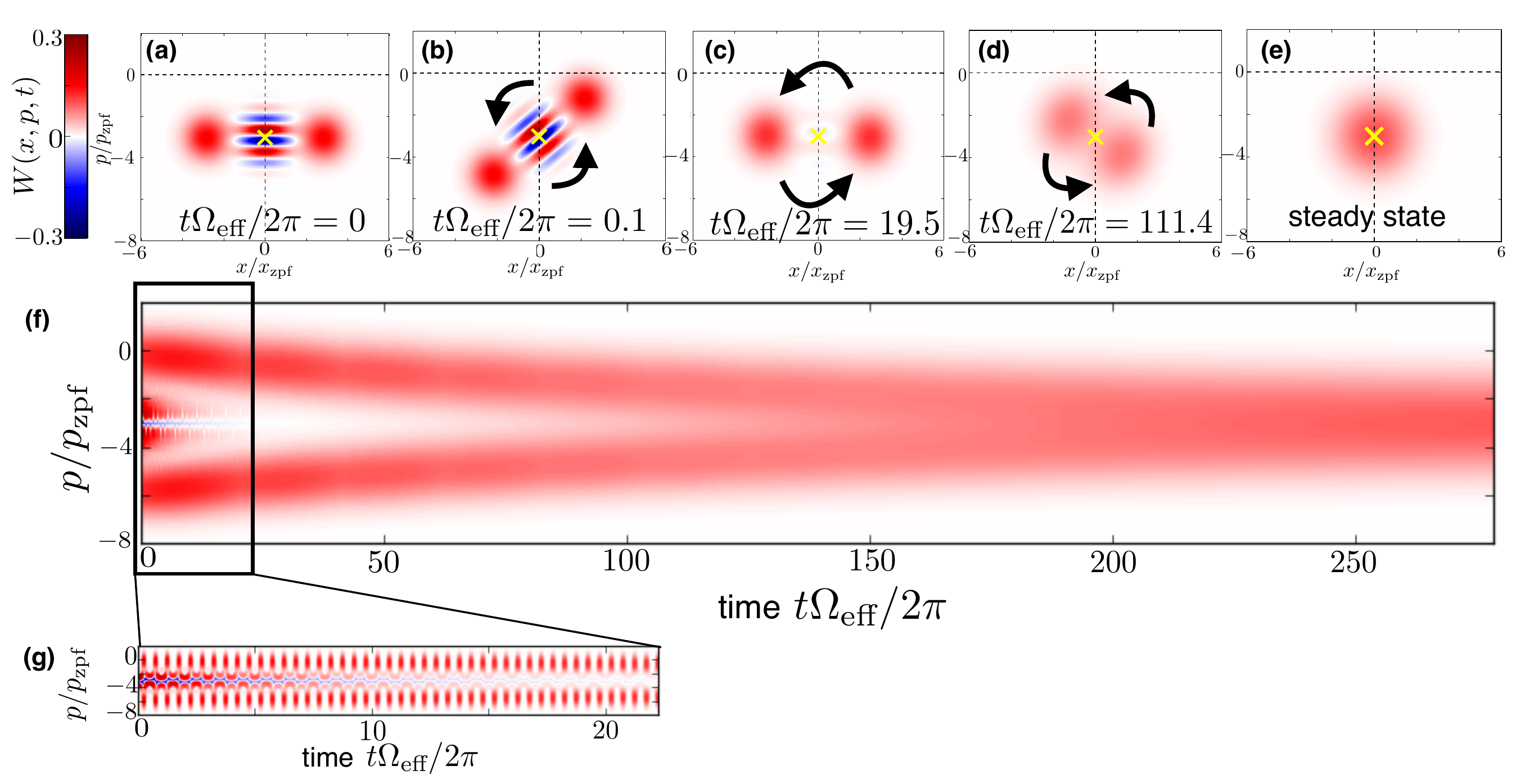}
	\caption{(color online).
	\emph{Long time evolution of coherent synchronization dynamics.}
	Wigner densities $W(x,p,t)$ of (a) an initial superposition state $\left|\Psi(t=0)\right\rangle \sim\left(\left|\beta_{ss}+2\right\rangle +\left|\beta_{ss}-2\right\rangle \right)$, (b)-(d) several snapshots at later times, and (e) the final steady state. The underdamped synchronization phase dynamics rotates the state around the classical steady-state solution (yellow cross). The interference fringes vanish due to dephasing and a classical mixture of displaced states remains, still rotating around the classical steady state. The displaced states eventually merge to the steady state on a timescale set by the damping.
	(e) and (f) show cuts of the Wigner densities, i.e.~$W(p,x=0, t)$, as a function of time. These time evolutions clearly show that dephasing and relaxing into the steady state occur on separate timescales. Parameters as in Fig.~3(a)-(c) of the main text: $\gamma_2/\gamma_1=0.1$, $F/\gamma_1=1.5\times 10^3$, and $\Delta/\gamma_1=7\times 10^2$.
	}
	\label{fig_LongTimeEvo}
\end{center}
\end{figure}

In the section 'Quantum coherence' in the main text we discuss that the synchronization dynamics can preserve quantum coherence for a significant number of oscillations of the system. 
Here, in Fig.~\ref{fig_LongTimeEvo}, we want to show how an initially prepared superposition state loses coherence and finally relaxes to the synchronized steady state. We numerically simulate the full quantum model to stress that this behaviour, expected due to a sufficiently small dephasing rate obtained from our effective quantum model, can indeed be observed (although quantitative deviations occur). The beginning of this time evolution is also shown and described in Fig.~3(a)-(c) of the main text, but will be repeated here for completeness. 

Starting with a superposition state, Fig.~\ref{fig_LongTimeEvo}(a), the Wigner density shows interference fringes with negativities. The synchronization dynamics described by the full master equation (1) of the main text, leads to rotations around the classical steady state (yellow cross), Fig.~\ref{fig_LongTimeEvo}(b). The frequency of these oscillations is, approximately, given by $\Omega_\text{eff}$ determined from the effective model. Due to dephasing, the interference fringes start to fade out, Fig.~\ref{fig_LongTimeEvo}(c), (f) and (g), and disappear after many oscillations of the system. Fig.~\ref{fig_LongTimeEvo}(c) shows a snapshot after almost $20$ oscillations, where the clear interference fringes have disappeared and only a small area of  slightly negative Wigner density values remains. However, the system is still far from its steady state, because the timescale set by the dephasing can be vastly different from the timescale set by the damping rate. In the example shown here, the state becomes a classical mixture of two displaced states first. Those displaced states merge at much later times,  Fig.~\ref{fig_LongTimeEvo}(d), and finally form the synchronized steady state of the system,  Fig.~\ref{fig_LongTimeEvo}(e).  Fig.~\ref{fig_LongTimeEvo}(f) and (g) show cuts along the momentum axis of the Wigner densities as a function of time. The damping, i.e.~the relaxing towards the steady state, can be viewed best in  the long time evolution Fig.~\ref{fig_LongTimeEvo}(f). To clearly see the dephasing, i.e.~the loss of coherence in form of vanishing negativities in the Wigner density,  Fig.~\ref{fig_LongTimeEvo}(g) zooms into the first part of the long time evolution. 

%

\section{Analytical spectrum}

In the over- and underdamped regime we can also obtain the spectrum 
from the analytical solution to our effective model. We start from Eq.~(4) of the main text, the squeezing Hamiltonian, and write down the quantum Langevine equations,

\begin{align}\label{qLeq1}
	\delta \dot {\hat b}&= i\Delta\delta\hat b-\frac{\Gamma}{2}\delta\hat b-\frac{\gamma_2}{2}\beta_{ss}^2\delta\hat b^\dagger+\sqrt{\Gamma}\hat\xi \, , \\
	\delta \dot {\hat b}^\dagger&= -i\Delta\delta\hat b^\dagger-\frac{\Gamma}{2}\delta \hat b^\dagger-\frac{\gamma_2}{2}\beta_{ss}^{*2}\delta\hat b+\sqrt{\Gamma}\hat\xi^\dagger.\label{qLeq2}
\end{align}

Here the noise operators $\hat\xi$ and $\hat\xi^\dagger$ represent white noise, fulfilling $\langle\hat\xi^\dagger(t)\hat\xi(t')\rangle=\bar n\delta(t-t')$ and $\langle\hat\xi(t)\hat\xi^\dagger(t')\rangle=(\bar n+1)\delta(t-t')$ and $\bar n=1/(4\gamma_2|\beta_{ss}|^2/\gamma_1-1)$ is obtained from the dissipation rates of Eq.~(3) in the main text, i.e. we identified $\gamma_1\equiv\bar n\Gamma$ and $4\gamma_2|\beta_{ss}|^2\equiv(\bar n+1)\Gamma$. Eqs.~(\ref{qLeq1}) and (\ref{qLeq2}) are easily solved in Fourier space where the problem simplifies to finding the inverse of a $2\times2$-matrix. Choosing the convention $\delta\hat b(\omega)=\int_{-\infty}^{+\infty} dt e^{i\omega t}\delta \hat b(t)$ and $\delta\hat b^\dagger(\omega)=\int_{-\infty}^{+\infty} dt e^{-i\omega t}\delta \hat b^\dagger(t)$ we find 

\begin{align*}
	\delta\hat b(\omega)&=\frac{\left[-i(\omega-\Delta)+\Gamma/2\right]\sqrt{\Gamma}}{\Delta^2-(\omega+i\Gamma/2)^2-\gamma_2^2|\beta_{ss}|^4}\hat\xi(\omega)-\frac{-\gamma_2\beta_{ss}^2\sqrt{\Gamma}}{\Delta^2-(\omega+i\Gamma/2)^2-\gamma_2^2|\beta_{ss}|^4}\hat\xi^\dagger(-\omega) \, , \\
	\delta\hat b^\dagger(\omega)&=-\frac{-\gamma_2\beta_{ss}^{*2}\sqrt{\Gamma}}{\Delta^2-(\omega+i\Gamma/2)^2-\gamma_2^2|\beta_{ss}|^4}\hat\xi(\omega)+\frac{\left[-i(\omega+\Delta)+\Gamma/2\right]\sqrt{\Gamma}}{\Delta^2-(\omega+i\Gamma/2)^2-\gamma_2^2|\beta_{ss}|^4}\hat\xi^\dagger(-\omega).
\end{align*}
Within the effective model the fluctuation spectrum $S_\trm{eff}(\omega)=\int_{-\infty}^{+\infty}dt e^{i\omega t}\langle\delta\hat b^\dagger(t)\delta \hat b(0)\rangle=\int_{-\infty}^{+\infty}\frac{d\omega'}{2\pi}\langle\delta\hat b^\dagger(-\omega)\delta\hat b(\omega')\rangle$ can be obtained from this solution by evaluating the relevant noise correlators. We find

\begin{align}
	S_\trm{eff}(\omega)=\frac{\Gamma\gamma_2^2|\beta_{ss}|^4+\bar n\Gamma\left[(\Gamma/2)^2+\gamma_2^2|\beta_{ss}|^4+(\omega+\Delta)^2\right]}{\left[(\omega-\sqrt{\Delta^2-\gamma^2|\beta_{ss}|^4})^2+(\Gamma/2)^2\right]\left[(\omega+\sqrt{\Delta^2-\gamma^2|\beta_{ss}|^4})^2+(\Gamma/2)^2\right]}.
\end{align}
This spectrum features peaks close to $\omega=\pm\sqrt{\Delta^2-\gamma_2^2|\beta_{ss}|^4}\equiv\Omega_\trm{eff}$.

\begin{figure}[b]
\begin{center}
	\includegraphics[width=0.5\columnwidth]{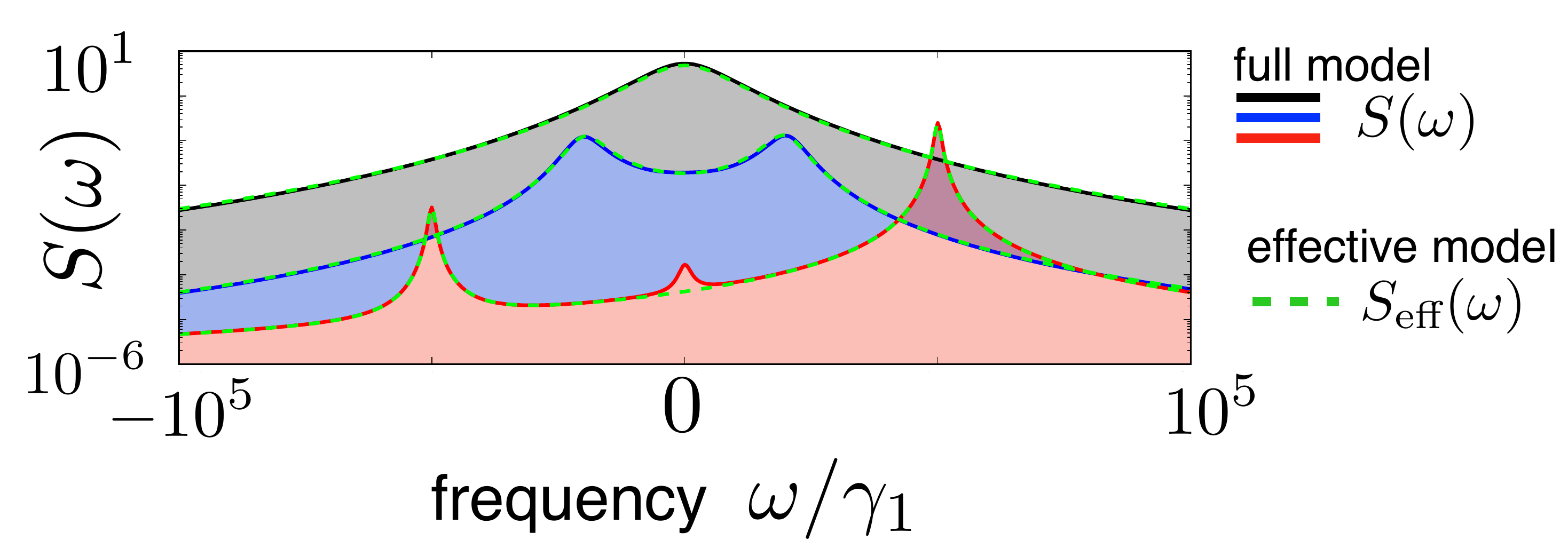}
	\caption{(color online).
	\emph{Full and effective spectrum.}
	The spectrum $S(\omega)$ of a synchronized VdP oscillator obtained from the full master equation (1) of the main text, and the corresponding $S_\trm{eff}(\omega)$ calculated 	analytically from the effective quantum model (dashed green lines). The three curves show the spectrum for different detuning $\Delta/\gamma_1=0$ (black), $\Delta/\gamma_1=2\times10^4$ (blue), and $\Delta/\gamma_1=5\times10^4$ (red). The effective spectrum features two (possibly degenerate) peaks close to $\pm\Omega_\trm{eff}$, and cannot capture the third peak appearing at larger detuning.
	Parameters as in Fig.~4(a) of the main text.
	}
	\label{fig_CompareSpectra}
\end{center}
\end{figure}

\bibliographystyle{apsrev4-1}

\begin{thebibliography}{99}

\bibitem{2001_Kurths_Synchronization}
J.\ Kurths, A.\ Pikovsky, and M.\ Rosenblum,
{\emph{Synchronization: A Universal Concept in Nonlinear Sciences}} (Cambridge University Press, 2001).

\bibitem{2013_Ludwig}
M.\ Ludwig and F.\ Marquardt,
\href{\doibase 10.1103/PhysRevLett.111.073603} {Phys.\ Rev.\ Lett.\ {\bf{111}}, 073603 (2013).}

\bibitem{2016_Weiss_OMsync_noiseInducedEffects}
T.\ Weiss, A.\ Kronwald, and F.\ Marquardt,
\href{\doibase 10.1088/1367-2630/18/1/013043}{New J.\ Phys.\ {\bf{18}}, 013043 (2016).}

\bibitem{2014_Minghui_QuantumSync_Atoms}
M.\ Xu, D.\ A.\ Tieri, E.\ C.\ Fine, J.\ K.\ Thompson, and M.\ J.\ Holland,
\href{\doibase 10.1103/PhysRevLett.113.154101}{Phys.\ Rev.\ Lett.\ {\bf{113}}, 154101 (2014).}

\bibitem{2014_Hush_SpinCorr_QSync}
M.\ R.\ Hush, W.\ Li, S.\ Genway, I.\ Lesanovsky, and A.\ D.\ Armour,
\href{\doibase 10.1103/PhysRevA.91.061401}{Phys.\ Rev.\ A {\bf{91}}, 061401 (2015).}

\bibitem{2013_Lee_QuantumSync_VdP_Ions}
T.\ E.\ Lee and H.\ R.\ Sadeghpour,
\href{\doibase 10.1103/PhysRevLett.111.234101}{Phys.\ Rev.\ Lett.\ {\bf{111}}, 234101 (2013).}

\bibitem{2014_Walter_QuantumSync_1Vdp}
S.\ Walter, A.\ Nunnenkamp, and C.\ Bruder,
\href{\doibase 10.1103/PhysRevLett.112.094102}{Phys.\ Rev.\ Lett.\ {\bf{112}}, 094102 (2014).}

\bibitem{2014_Lee_EntanglementTongue_QSync}
T.\ E.\ Lee, C.\ K.\ Chan, and S.\ Wang,
\href{\doibase 10.1103/PhysRevE.89.022913}{Phys.\ Rev.\ E {\bf{89}}, 022913 (2014).}

\bibitem{2014_Walter_SynchronizationVdPosc}
S.\ Walter, A.\ Nunnenkamp, and C.\ Bruder,
\href{\doibase 10.1002/andp.201400144}{Ann.\ Phys.\ (Berlin) {\bf{527}}, 131 (2014).}

\bibitem{2016_Loerch_quantumSync_vdPKerr}
N.\ L{\"o}rch, E.\ Amitai, A.\ Nunnenkamp, and C.\ Bruder,
\href{\doibase 10.1103/PhysRevLett.117.073601}{Phys.\ Rev.\ Lett.\ {\bf{117}}, 073601 (2016).}

\bibitem{2008_Vinokur_Qsync} 
V.\ M.\ Vinokur, T.\ I.\ Baturina, M.\ V.\ Fistul, A.\ Y.\ Mironov, M.\ R.\ Baklanov, and C.\ Strunk,
\href{\doibase 10.1038/nature06837}{Nature {\bf{452}}, 613 (2008).}

\bibitem{2013_Hriscu_QSync_SuperconductingDevice}
A.\ M.\ Hriscu and Y.\ V.\ Nazarov,
\href{\doibase 10.1103/PhysRevLett.110.097002}{Phys.\ Rev.\ Lett.\ {\bf{110}}, 097002 (2013).}

\bibitem{2013_Mari_QuantumSync}
A.\ Mari, A.\ Farace, N.\ Didier, V.\ Giovannetti, and R.\ Fazio,
\href{\doibase 10.1103/PhysRevLett.111.103605}{Phys.\ Rev.\ Lett.\ {\bf{111}}, 103605 (2013).}

\bibitem{2015_Ameri_MutualInformation_QuantumSync}
V.\ Ameri, M.\ Eghbali-Arani, A.\ Mari, A.\ Farace, F.\ Kheirandish, V.\ Giovannetti, and R.\ Fazio,
\href{\doibase 10.1103/PhysRevLett.111.103605}{Phys.\ Rev.\ A {\bf{91}}, 012301 (2015).}

\bibitem{2005_Acebron_KuramotoReview}
J.\ A.\ Acebr{\'o}n, L.\ L.\ Bonilla, C.\ J.\ P{\'e}rez Vicente, F.\ Ritort, and R.\ Spigler,
\href{\doibase 10.1103/RevModPhys.77.137}{Rev.\ Mod.\ Phys.\ {\bf{77}}, 137 (2005).}

\bibitem{2012_ZhangLipson_SynchronizationPRL}
M.\ Zhang, G.\ S.\ Wiederhecker, S.\ Manipatruni, A.\ Barnard, P.\ McEuen, and M.\ Lipson,
\href{\doibase 10.1103/PhysRevLett.109.233906}{Phys.\ Rev.\ Lett.\ {\bf{109}}, 233906 (2012).}

\bibitem{2013_BagheriTang_Synchronization}
M.\ Bagheri, M.\ Poot, L.\ Fan, F.\ Marquardt, and H.\ X.\ Tang,
\href{\doibase 10.1103/PhysRevLett.111.213902}{Phys.\ Rev.\ Lett.\ {\bf{111}}, 213902 (2013).}

\bibitem{2015_Shlomi_SyncExtDrive}
K.\ Shlomi, D.\ Yuvaraj, I.\ Baskin, O.\ Suchoi, R.\ Winik, and E.\ Buks,
\href{\doibase 10.1103/PhysRevE.91.032910}{Phys.\ Rev.\ E {\bf{91}}, 032910 (2015).}

\bibitem{2015_Lipson_ArraySync}
M.\ Zhang, S.\ Shah, J.\ Cardenas, and M.\ Lipson,
\href{\doibase 10.1103/PhysRevLett.115.163902}{Phys.\ Rev.\ Lett.\ {\bf{115}}, 163902 (2015).}

\bibitem{2011_Heinrich_CollectiveDynamics}
G.\ Heinrich, M.\ Ludwig, J.\ Qian, B.\ Kubala, and F.\ Marquardt,
\href{\doibase 10.1103/PhysRevLett.107.043603}{Phys.\ Rev.\ Lett.\ {\bf{107}}, 043603 (2011).}

\bibitem{2014_Lauter_PhasePatterns}
R.\ Lauter, C.\ Brendel, S.\ J.\ M.\ Habraken, and F.\ Marquardt,
\href{\doibase 10.1103/PhysRevE.92.012902}{Phys.\ Rev.\ E {\bf{92}}, 012902 (2015).}

\bibitem{1988_Chakraborty_VdP_PhaseSync}
T.\ Chakraborty and R.\ H.\ Rand,
\href{\doibase 10.1016/0020-7462(88)90034-0}{Int.\ J.\ Non Linear Mech.\ {23} 369 (1988).}

\bibitem{1990_Aronson_VdP_PhaseSync}
D.\ G.\ Aronson, G.\ B.\ Ermentrout, and N.\ Kopell,
\href{\doibase 10.1016/0167-2789(90)90007-C}{Physica D {41} 403 (1990).}

\bibitem{2000_Pikovsky_PhaseSync} 
A.\ Pikovsky, M.\ Rosenblum, and J.\ Kurths,
Int.\ J.\ Bifurcation Chaos {\bf{10}}, 2291 (2000).

\bibitem{2014_Barois_SOSO}
T.\ Barois, S.\ Perisanu, P.\ Vincent, S.\ T.\ Purcel, and A.\ Ayari,
\href{\doibase 10.1088/1367-2630/16/8/083009}{New J.\ Phys.\ {\bf{16}}, 083009 (2014).}

\bibitem{2011_Thevenin_PhaseTrapping_Laser}
J.\ Th\'evenin, M.\ Romanelli, M.\ Vallet, M.\ Brunel, and T.\ Erneux,
\href{\doibase 10.1103/PhysRevLett.107.104101}{Phys.\ Rev.\ Lett.\ {\bf{107}}, 104101 (2011).}

\bibitem{2005_Trees_SyncJosephsonJunctions}
B.\ R.\ Trees, V.\ Saranathan, and D.\ Stroud,
\href{\doibase 10.1103/PhysRevE.71.016215}{Phys.\ Rev.\ E {\bf{71}}, 016215 (2005).}

\bibitem{supp2016}
See Supplemental Material for details.

\bibitem{2008_WallsMilburn_QuantumOptics}
D.\ Walls and G.\ J.\ Milburn,
{\emph{Quantum Optics}} (Springer-Verlag Berlin Heidelberg, 2008).

\bibitem{footnote1}
Quantum noise leads to a finite threshold for synchronization~\cite{2013_Ludwig, 2014_Walter_QuantumSync_1Vdp}. Below this threshold, the effective model is not applicable since classically a stable fixed point exists, but the quantum system settles on a LC. To improve the readability of 2(b) we nevertheless also plot the region where the effective model fails which here is only the case for very small force and detuning.

\bibitem{2011_Dykman_QuantumHeating}
M.\ I.\ Dykman, M.\ Marthaler, and V.\ Peano,
\href{\doibase 10.1103/PhysRevA.83.052115}{Phys.\ Rev.\ A {\bf{83}}, 052115 (2011).}

\bibitem{1996_Leibfried_QuantumStatesTrappedIons}
D.\ Leibfried, D.\ M.\ Meekhof, B.\ E.\ King, C.\ Monroe, W.\ M.\ Itano, and D.\ J.\ Wineland,
\href{\doibase 10.1103/PhysRevLett.77.4281}{Phys.\ Rev.\ Lett.\ {\bf{77}}, 4281 (1996).}

\bibitem{2012_Islam_PhDThesis}
K.\ R.\ Islam, Ph.D.\ thesis, University of Maryland (2012)

\bibitem{2003_Leibfried_RevMod_Ions}
D.\ Leibfried, R.\ Blatt, C.\ Monroe, and D.\ Wineland,
\href{\doibase 10.1103/RevModPhys.75.281}{Rev.\ Mod.\ Phys.\ {\bf{75}}, 281 (2003).}

\bibitem{2006_FM_DynamicalMultistability}
F.\ Marquardt, J.\ G.\ E.\ Harris, and S.\ M.\ Girvin,
\href{\doibase 10.1103/PhysRevLett.96.103901}{Phys.\ Rev.\ Lett.\ {\bf{96}}, 103901 (2006).}

\bibitem{2005_KippenbergVahala_SelfOsc}
T.\ J.\ Kippenberg, H.\ Rokhsari, T.\ Carmon, A.\ Scherer, and K.\ J.\ Vahala,
\href{\doibase 10.1103/PhysRevLett.95.033901}{Phys.\ Rev.\ Lett.\ {\bf{95}}, 033901 (2005).}

\bibitem{2008_Metzger_SelfOsc}
C.\ Metzger, M.\ Ludwig, C.\ Neuenhahn, A.\ Ortlieb, I.\ Favero, K.\ Karrai, and F.\ Marquardt,
\href{\doibase 10.1103/PhysRevLett.101.133903}{Phys.\ Rev.\ Lett.\ {\bf{101}}, 133903 (2008).}

\bibitem{2008_Hofheinz_GeneratingFockStates}
M.\ Hofheinz, E.\ M.\ Weig, M.\ Ansmann, R.\ C.\ Bialczak, E.\ Lucero, M.\ Neeley, A.\ D.\ O'Connell, H.\ Wang, J.\ M.\ Martinis, and A.\ N.\ Cleland,
\href{\doibase 10.1038/nature07136}{Nature {\bf{454}}, 310 (2008).}

\bibitem{2008_Deleglise_EngineeringQuantumStates}
S.\ Deleglise, I.\ Dotsenko, C.\ Sayrin, J.\ Bernu, M.\ Brune, J.\-M.\ Raimond, and S.\ Haroche,
\href{\doibase 10.1038/nature07288}{Nature {\bf{455}}, 510 (2008).}

\bibitem{2009_Hofheinz_ArbitraryQuantumStates}
M.\ Hofheinz, H.\ Wang, M.\ Ansmann, R.\ C.\ Bialczak, E.\ Lucero, M.\ Neeley, A.\ D.\ O'Connell, D.\ Sank, J.\ Wenner, J.\ M.\ Martinis, and A.\ N.\ Cleland,
\href{\doibase 10.1038/nature08005}{Nature {\bf{459}}, 546 (2009).}

\bibitem{2015_Leghtas_EngineeredTwoPhotonLoss}
Z.\ Leghtas, S.\ Touzard, I.\ M.\ Pop, A.\ Kou, B.\ Vlastakis, A.\ Petrenko, K.\ M.\ Sliwa, A.\ Narla, S.\ Shankar, M.\ J.\ Hatridge, M.\ Reagor, L.\ Frunzio, R.\ J.\ Schoelkopf, M.\ Mirrahimi, and M.\ H.\ Devoret,
\href{\doibase 10.1126/science.aaa2085}{Science {\bf{347}}, 853 (2015).}

\end{thebibliography}

\begin{thebibliography}{99}

\bibitem{2001_Kurths_Synchronization}
J.\ Kurths, A.\ Pikovsky, and M.\ Rosenblum,
{\emph{Synchronization: A Universal Concept in Nonlinear Sciences}} (Cambridge University Press, 2001).

\bibitem{2010_Broer_HandbookDynSys}
H.~W.\ Broer, B.\ Hasselblatt, and F.\ Takens,
{\emph{Handbook of Dynamical Systems - Volume 3}} (North Holland, 2010).

\end{thebibliography}


\end{document}